\documentclass[12pt]{article}

\usepackage{sbc-template}
\usepackage{threeparttable}
\usepackage{graphicx,url}

\usepackage{array}
\usepackage[brazil]{babel}   

\title{Conformidade com os Requisitos Legais de Privacidade de Dados: Um Estudo sobre Técnicas de Anonimização}

\author{André Menolli\inst{1}\inst{2}, Luiz Fernando  Nunes\inst{2}, Thiago A. Coleti\inst{1} }

\address{Centro de Ciências Tecnológicas -- Universidade Estadual do Norte do Paraná  (UENP)\\
  Paraná, Brazil
\nextinstitute
 Programa de Pós-Graduação em Ciência da Computação\\ Unversidade Estadual de Londrina (UEL)--
 Paraná, Brazil
 \email{\{menolli,thiago.coleti\}@uenp.edu.br, luiz.fernando.pereira@uel.br}
}

\begin{document} 

\maketitle

\begin{abstract}
 The protection of personal data has become a central topic in software development, especially with the implementation of the General Data Protection Law (LGPD) in Brazil and the General Data Protection Regulation (GDPR) in the European Union. With the enforcement of these laws, certain software quality criteria have become mandatory, such as data anonymization, which is one of the main aspects addressed by these regulations. The aim of this article is to analyze data anonymization techniques and assess their effectiveness in ensuring compliance with legal requirements and the utility of the data for its intended purpose. Techniques such as aggregation, generalization, perturbation, and k-anonymity were investigated and applied to datasets containing personal and sensitive data. The analysis revealed significant variations in the effectiveness of each method, highlighting the need to balance privacy and data utility.
\end{abstract}
     
\begin{resumo} 
A proteção de dados pessoais tornou-se um tema central no desenvolvimento de software, especialmente com a implementação da Lei Geral de Proteção de Dados (LGPD) no Brasil e do Regulamento Geral de Proteção de Dados (GDPR) na União Europeia. Com a entrada em vigor dessas legislações, certos critérios, como a privacidade dos dados, tornaram-se aspectos fundamentais dessas normas. Este artigo tem como objetivo implementar uma abordagem computacional para anonimização, utilizando diferentes técnicas, e analisar seus resultados, além de investigar como essas técnicas se alinham aos requisitos da LGPD. Foram exploradas e aplicadas estratégias como agregação, generalização, perturbação e k-anonimato a conjuntos de dados contendo informações pessoais e sensíveis. A análise revelou variações significativas na eficácia de cada método, destacando a necessidade de equilibrar privacidade e utilidade dos dados.
\end{resumo}

\section{Introdução} \label{sec:intro}

A proteção de dados pessoais tornou-se um tema central nos sistemas de informação nos últimos anos, especialmente com a implementação da Lei Geral de Proteção de Dados (LGPD) \cite{LGPD2018} no Brasil e do Regulamento Geral sobre a Proteção de Dados (GDPR) \cite{GDPR2016} na União Europeia. Ambos os regulamentos têm como objetivo garantir a privacidade e a segurança dos dados pessoais, impactando diretamente a maneira como o software é projetado, desenvolvido e mantido.\par
Dados pessoais referem-se a qualquer informação relacionada a uma pessoa natural identificada ou identificável, incluindo categorias especiais de dados sensíveis, como informações sobre saúde, orientação sexual e convicções religiosas \cite{LGPD2018}. A LGPD e o GDPR exigem uma gestão rigorosa desses dados, na qual o controle e a transparência são fundamentais. Para demonstrar conformidade, as empresas devem documentar detalhadamente as atividades de processamento de dados e implementar políticas e procedimentos claros para sua proteção. Isso inclui a adoção dos princípios de privacidade desde a concepção e por padrão (\textit{Privacy by Design e Privacy by Default}), garantindo que os sistemas sejam desenvolvidos para proteger os dados pessoais de forma eficaz \cite{Maldonado2020}. \par
Um dos direitos fundamentais assegurados tanto pela LGPD quanto pelo GDPR é o direito à portabilidade dos dados \cite{LGPD2018, GDPR2016}. Esse direito permite que os titulares dos dados obtenham suas informações pessoais em um formato estruturado, de uso comum e legível por máquina, possibilitando sua transferência para outro controlador sem impedimentos. A portabilidade promove maior transparência e empodera os titulares ao oferecer mais controle sobre seus próprios dados \cite{Kateifides2020}.\par
Além disso, a conformidade com a LGPD e o GDPR exige que as empresas revisem seus contratos com terceiros que processam dados em seu nome \cite{LGPD2018, GDPR2016}. Esses contratos devem assegurar que os prestadores de serviço cumpram os requisitos de proteção de dados, incluindo cláusulas específicas sobre segurança e privacidade \cite{Kateifides2020}.
A implementação de medidas técnicas e organizacionais adequadas é essencial para garantir a conformidade com ambos os regulamentos. Essas medidas incluem criptografia de dados, controles de acesso rigorosos, auditorias regulares de segurança e políticas de segurança da informação \cite{Martin2020}. Além disso, é fundamental estabelecer diretrizes claras para a coleta, armazenamento, uso, compartilhamento e descarte de dados pessoais, assegurando que todas as práticas estejam alinhadas às exigências legais \cite{Maldonado2020}. 
Uma estratégia que pode ser utilizada para se manter a privacidade dos dados e atender às exigências da LGPD e do GDPR é a anonimização de dados. Essa técnica visa tornar irreconhecível o vínculo entre os dados e seus titulares, reduzindo os riscos associados ao tratamento de informações pessoais. No entanto, a eficácia da anonimização depende das características dos dados e das técnicas empregadas, uma vez que diferentes métodos podem oferecer níveis variados de proteção. Além disso, os avanços tecnológicos têm viabilizado o desenvolvimento de técnicas que, em alguns casos, permitem reverter a anonimização, comprometendo a privacidade dos dados. \par

Alguns estudos, como \cite{rana2016privacy}, investigam como a anonimização pode contribuir para a preservação da privacidade dos dados. Já \cite{marques2020analysis} analisam diversas técnicas para avaliar sua eficácia na proteção de dados conforme os requisitos da GDPR. Um estudo mais recente sobre anonimização é o de \cite{sainz2022python}, que apresenta a biblioteca Python pyCANON. Essa ferramenta permite a avaliação do nível de anonimato de conjuntos de dados, aplicando múltiplas técnicas para garantir maior segurança e conformidade com padrões regulatórios.

No entanto, a literatura ainda carece de estudos que abordem métodos de anonimização automatizados e analisem sua conformidade com as legislações de proteção de dados, especialmente a LGPD. Diante disso, este artigo apresenta uma abordagem computacional utilizando quatro das principais técnicas de anonimização e analisa os resultados de sua aplicação, especialmente no atendimento aos requisitos da LGPD. Para isso, foi conduzido um experimento no qual um conjunto de dados foi anonimizado, permitindo a avaliação da qualidade e do nível de privacidade das informações resultantes das técnicas aplicadas. Como principais objetivos desse trabalho destacam-se:

\begin{enumerate}
    \item Analisar comparativamente as técnicas de anonimização quanto à sua eficácia em garantir privacidade e preservar a utilidade dos dados;
    \item Implementar um método computacional para anonimizar as técnicas selecionadas;
    \item Avaliar as técnicas de anonimização por meio da aplicação do método computacional proposto.
\end{enumerate}
 \par

\section{Trabalhos Relacionados} \label{sec:firstpage}

A privacidade de dados tornou-se um desafio crítico com a crescente digitalização em diversos setores. A anonimização emerge como uma estratégia fundamental para proteger informações pessoais, permitindo análises sem riscos significativos de exposição. Nesse contexto, \cite{marques2020analysis} exploram a generalização como uma abordagem eficaz para preservar a privacidade sem comprometer a qualidade das análises. Além disso, \cite{Ramos2019} e \cite{RamosOliveira2018} discutem a aplicabilidade prática da anonimização em pesquisa e negócios, destacando sua importância para alinhar a proteção de dados às exigências regulatórias. Entre as técnicas avançadas, \cite{LiLi2014} propõem o \textit{slicing}, um método que busca equilibrar privacidade e utilidade dos dados.\par
Do ponto de vista legal, \cite{Senigaglia2020} enfatizam as implicações regulatórias e a necessidade de conformidade com múltiplas jurisdições. Já \cite{RamosOliveira2020} apontam que as técnicas de anonimização devem evoluir continuamente para enfrentar desafios emergentes e garantir análises seguras e eficazes. No entanto, estudos indicam que métodos tradicionais, como k-anonimato e l-diversidade, podem ser vulneráveis a ataques avançados, reforçando a necessidade de maior automação e métricas robustas para avaliar sua eficácia. \par

 Dentre os trabalhos existentes sobre anonimização, destaca-se o de \cite{ranjan2016two}, que avalia técnicas como k-anonimato e l-diversidade, identificando vulnerabilidades nessas abordagens tradicionais, especialmente diante de ataques de reidentificação em conjuntos de dados complexos. Apesar de sua contribuição ao expor essas fragilidades, o estudo não aborda métodos para automatizar a avaliação da qualidade dos dados anonimizados.\par
A introdução do conceito de k-anonimato por \cite{samarati1998protecting} representou um marco na área, garantindo que cada registro seja indistinguível de outros k-1 registros. No entanto, essa técnica apresenta limitações importantes, como a dificuldade de adaptação a grandes volumes de dados e a incapacidade de lidar com correlações complexas, restringindo sua aplicabilidade em cenários mais avançados.\par
Em uma revisão abrangente, \cite{mogre2012review} analisaram técnicas como generalização e perturbação, ressaltando a importância da escolha adequada de métodos conforme o contexto dos dados. Entretanto, o estudo dá pouca ênfase à preservação da utilidade dos dados anonimizados, aspecto essencial para aplicações práticas em estatística e aprendizado de máquina.\par 

O trabalho de \cite{SmithChang2021} investigaram a adaptação de técnicas de anonimização para\textit{ big data} e aprendizado de máquina, trazendo contribuições relevantes para contextos modernos. No entanto, uma limitação notável é a ausência de automação nos testes e nas métricas de avaliação, fatores essenciais para garantir a eficácia das técnicas em cenários dinâmicos.
O trabalho de \cite{rana2016privacy} analisa a importância de preservar a privacidade sem comprometer a utilidade dos dados na área da saúde, conduzindo um estudo comparativo de técnicas e avaliando ameaças potenciais.

\par No contexto das regulamentações, \cite{marques2020analysis} exploram os conceitos de anonimização e pseudonimização, examinando técnicas e ferramentas para determinar quais oferecem maior nível de proteção, bem como suas vantagens e limitações. No entanto, este estudo se concentra na GDPR e apresenta lacunas em relação à automação de testes e métricas de avaliação. \par

\section{Método de Pesquisa}

O método de pesquisa adotado teve como objetivo analisar diferentes técnicas de anonimização aplicadas a dois datasets distintos, que diferem tanto na quantidade de campos quanto no contexto. O foco da análise foi compreender o comportamento dessas técnicas em conjuntos de dados que podem ou não conter informações sensíveis.

\begin{figure}[ht]
\centering
\includegraphics[width=.6\textwidth]{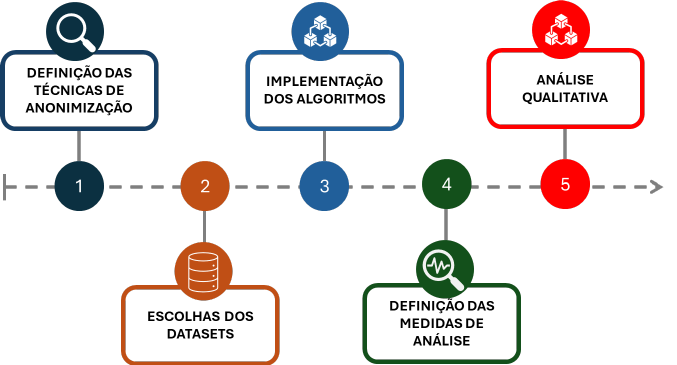}
\caption{Fluxo do Método de Pesquisa aplicado no Estudo}
\label{fig:method}
\end{figure}

A Figura \ref{fig:method} apresenta o método de pesquisa conduzido, estruturado em cinco etapas. Cada uma dessas etapas é detalhada nas subseções seguintes.

\subsection{Definição das Técnicas de Anonimização}

As técnicas de anonimização são frequentemente utilizadas para preservar a privacidade \cite{muntes2009privacy}, com foco na conversão de dados pessoais em dados anonimizados para reduzir o risco de divulgação não autorizada. Independentemente do método empregado, espera-se que essas técnicas reduzam, em alguma medida, as informações originais contidas no conjunto de dados.

Neste trabalho, foram escolhidas quatro técnicas de anonimização para avaliar sua efetividade nesse processo e verificar se seu uso pode atender às exigências legais. A seleção dessas técnicas foi baseada em três critérios principais:

\begin{enumerate}
\item Relevância legal: todas as técnicas escolhidas são recomendadas para conformidade com a LGPD e o GDPR.
\item Efetividade: essas técnicas são reconhecidas por oferecer um equilíbrio entre proteção da privacidade e preservação da utilidade dos dados.
\item Facilidade de aplicação prática: são métodos que podem ser implementados com ferramentas de código aberto e amplamente disponíveis, facilitando a reprodutibilidade dos experimentos.
\end{enumerate}


Dentre as técnicas selecionadas, a agregação consolida informações em grupos para dificultar a identificação individual, sendo destacada por sua simplicidade e utilidade prática. Já a generalização, que substitui valores específicos por intervalos ou categorias mais amplas, é reconhecida por preservar a utilidade dos dados em níveis aceitáveis \cite{ciriani2009theory}.

A perturbação, que modifica ligeiramente os dados reais para preservar a privacidade, é valorizada por sua capacidade de equilibrar proteção e utilidade \cite{liew1985data}, especialmente em análises estatísticas \cite{rana2016privacy}. Por fim, o k-anonimato, técnica pioneira introduzida por Samarati e Sweeney, assegura que cada registro seja indistinguível de pelo menos k-1 outros, sendo amplamente adotado, embora apresente limitações em conjuntos de dados com atributos altamente correlacionados \cite{samarati1998protecting}.

\subsection{Escolha dos \textit{Datasets} e Implementação do Método Computacional}
Cada técnica foi aplicada a dois \textit{datasets} distintos, ambos compostos por 250 registros sintéticos. O primeiro dataset contém apenas dados pessoais, enqaunto o segundo \textit{dataset}, além dos dados pessoais, inclui dados sensíveis. Isso permite uma análise mais detalhada das técnicas de anonimização em cenários que apresentam maior risco de reidentificação.\par

A implementação computacional das técnicas de anonimização de dados foi realizada utilizando bibliotecas e funções específicas do \textit{Python}, como \textit{pandas} e \textit{numpy}.

A técnica de agregação foi implementada com a biblioteca \textit{pandas}, agrupando valores individuais em categorias maiores para reduzir a granularidade dos dados. Já a técnica de generalização substituiu valores específicos por categorias mais amplas, com o objetivo de reduzir a especificidade dos dados. 
A técnica de perturbação introduziu ruído nos dados para dificultar a identificação direta dos indivíduos. Em nossa abordagem, foi adicionado ruído gaussiano, garantindo a proteção da privacidade enquanto preserva a utilidade estatística dos dados.

Por fim, a técnica de k-anonimato foi implementada para assegurar que cada registro fosse indistinguível de pelo menos k-1 outros registros em relação a determinados atributos. Para isso, foram utilizadas estruturas de dados do \textit{pandas}, como \textit{DataFrames}, e contadores de frequência, permitindo identificar registros únicos ou raros.

\subsection{Definição das Medidas de Análise}
As métricas de entropia e perda percentual de informação foram utilizadas para avaliar a eficácia dos algoritmos de anonimização. A entropia mede a imprevisibilidade dos dados, refletindo a dificuldade de reidentificação e, portanto, a privacidade dos indivíduos \cite{baez2011entropy, xu2007entropy}. Já a perda de informação avalia o impacto da anonimização na utilidade dos dados, garantindo um equilíbrio entre privacidade e relevância analítica \cite{ghinita2007anonymization}. Para isso, a informação útil foi definida com base nos atributos essenciais para as análises, e a perda foi quantificada comparando a entropia dos dados originais com a dos dados anonimizados.\par
O cálculo da perda de informação é essencial para entender o trade-off entre privacidade e utilidade. Uma alta perda indica que a anonimização compromete a qualidade dos dados, dificultando análises precisas, enquanto uma baixa perda sugere que os dados permanecem úteis sem comprometer excessivamente a privacidade. Dependendo do contexto, diferentes níveis de generalização ou perturbação podem ser mais adequados para garantir tanto a proteção da privacidade quanto a preservação da informação essencial.

\subsection{Análise Quantitativa}

A análise quantitava dos resultados foi realizada considerando as métricas definidas, com intuito de avaliar a perda de informação após a anonimização. A entropia compara a quantidade de informação nos dados originais e anonimizados, fornecendo uma medida clara do impacto da anonimização. 

\section{Resultados}\label{resultados}
Nesta seção, são apresentados os resultados das técnicas de anonimização aplicadas aos conjunto de dados, os quais contém dados pessoais e sensíveis. Conforme apresentado na Tabela \ref{tab:resultados}, são exibidos os resultados da entropia inicial dos dados, da entropia do dados anonimizados e da perda de informação após a aplicação das técnicas de anonimização. \par

\begin{flushleft}
\begin{table}[h]
\begin{threeparttable}
\centering

\scalebox{0.83}{
\begin{tabular}{|l|p{0.7cm}|p{0.7cm}|p{0.9cm}|p{0.7cm}|p{0.7cm}|p{0.9cm}|p{0.7cm}|p{0.7cm}|p{0.9cm}|p{0.7cm}|p{0.7cm}|p{0.9cm}|c|}
\hline
 & \multicolumn{12}{c|}{\textbf{Técnicas de Anonimização}} \\
\hline
\textbf{Campo}
 & \multicolumn{3}{c|}{\textbf{Generalização}} & \multicolumn{3}{c|}{\textbf{K-Anonimato}} & \multicolumn{3}{c|}{\textbf{Agregação}}  & \multicolumn{3}{c|}{\textbf{Perturbação}}

\\
\cline{2-13}
& Entr. Or. & Entr. An. & Perda (\%) & Entr. Or. & Entr. An. & Perda (\%) & Entr. Or. & Entr. An. & Perda (\%) & Entr. Or. & Entr. An. & Perda (\%) \\
\hline
Nome* & 7.8 & 7.4 & 4.9\% & 7.8 & 7.4 & 5.2\% & 7.8 & 7.4 & 4.9\% &7.8&7.8&0\% \\
CPF* & 7.8 & 7.7 & 0.46\% & 7.8 & 6.5 & 16.1\% & 7.8 & 6.5 & 16\% &7.8&7.6&2.1\%  \\
E-mail* & 7.8 & 1.4 & 82.4\% & 7.8 & 4.6 & 41.4\% & 7.8 & 4.5 & 41.5\% &7.8&7.8&0\%  \\
Endereço* & 7.8 & 4.3 & 45.1\% & 7.8 & 4.3 & 45.2\% & 7.8 & 4.3 & 45.1\% &7.8&7.8&0\% \\
Faixa Etária & 0.0 & 0.25 & 0\% & 0.0 & 1.1 & 0\% & 0.0 & 1.1 & 0\% &0 &4.2&0\%  \\
Cidade & 4.2 & 0.0 & 100\% & 4.2 & 0.0 & 100\%& 4.2 & 0.0 & 100\% &4.2&4.2&0\% \\
Estado & 0.4 & 0.0 & 100\% & 0.4 & 0.0 & 100\% & 0.4 & 0.0 & 100\% &&&\\
Estado Civil & 0.2 & 0.2 & 0\% & 0.2 & 0.1 & 45.5\% & 0.2 & 0.2 & 0\% &&&\\
Cor & 1.4 & 1.4 & 0\% & 1.4 & 1.3 & 5.2\% & 1.4 & 1.4 & 0\% &&&\\
Problema de Saúde &&&&&&&2.7&2.7&0\%&&&\\

\hline
\end{tabular}
}

\begin{tablenotes}
\small
\item * Dados Pessoais
\item Entr. Or - Entropia original
\item Entr. An. - Entropia após aplicar técnica de anonimização
\item Perda (\%) - percentual de quanto da informação original foi perdida devido à técnica aplicada

\end{tablenotes}
\end{threeparttable}
\caption{Resultados da Entropia Original e Anonimizada utilizando diferentes técnicas de anonimização aplicadas ao DataSet de Dados Pessoais e Sensíveis}
\label{tab:resultados}
\end{table}

\end{flushleft}

A a entropia original apresenta  a variabilidade dos dados antes da anonimização, enquanto a entropia anonimizada mede a variabilidade dos dados após a anonimização. A partir destes dois dados é possível entender qual foi a perda de informação, ou diferença entre a entropia original e a entropia após a anonimização, que na Tabela \ref{tab:resultados} é apresentada de forma percentual.

Cada uma das quatros técnicas aplicadas ao dataset possuem estratégias distintas para a anonimização dos dados, por isso a diferença entre os resutlados. Para entender um pouco melhor como cada técnica funciona, a Tabela \ref{tab:resultados_analise} sumariza a estratégia de anonimização de cada uma das técnicas para cada campo do dataset.

\begin{flushleft}
\begin{table}[h]
\scalebox{0.51}{
\begin{tabular}{c}
    \includegraphics{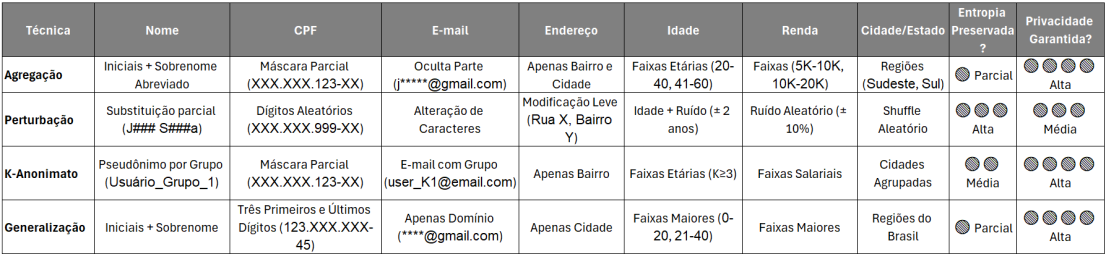}
\end{tabular}
}
\caption{Estratégias de Ananonimização aplicadas por cada uma das técnicas}
\label{tab:resultados_analise}
\end{table}
\end{flushleft}
Considerando os resultados apresentados na Tabela \ref{tab:resultados} e a visão geral de como cada técnica funciona (Tabela \ref{tab:resultados_analise}), nas próximas subseções é apresentada uma breve descrição dos resultados de cada técnica.

\subsection{Agregação}

A técnica de agregação, ao substituir valores individuais por categorias mais amplas, como grupos etários, regiões geográficas e faixas de renda, impacta de forma distinta os dados pessoais e os dados sensíveis.

No que diz respeito aos dados pessoais, como a faixa etária, a agregação resultou em uma entropia maior (1,069), o que pode indicar um erro de cálculo ou um efeito colateral do processo. Já atributos como cidade, estado e estado civil perderam completamente sua variabilidade, sendo reduzidos a um único valor, o que elimina detalhes individuais importantes. Além disso, dados pessoais como nome, CPF, e-mail e endereço foram completamente removidos do conjunto de dados, enquanto atributos detalhados, como cidades, foram substituídos por categorias amplas, como regiões. Essa abordagem garante máxima proteção à privacidade, pois os identificadores diretos não estão mais presentes. No entanto, isso resulta na perda total de individualidade, tornando inviável a realização de análises detalhadas que dependam de dados granulares.

Por outro lado, em relação aos dados sensíveis, como informações sobre problemas de saúde, a agregação também resultou na perda total da variabilidade, comprometendo a capacidade de identificar padrões ou tendências em grupos específicos. A substituição de atributos detalhados por categorias amplas reduz a precisão na identificação de padrões e tendências, o que pode limitar a utilidade do conjunto de dados para análises mais específicas. Assim, embora a privacidade seja garantida, a capacidade analítica é comprometida.

\subsection{Perturbação}

A técnica de perturbação, ao adicionar ruído aleatório aos dados, mantém os valores próximos aos originais, mas com algumas alterações. Em termos de entropia, os resultados indicam uma alteração mínima. A entropia não sofreu variação significativa para atributos como cidade, estado civil e cor, sugerindo que a perturbação não afetou consideravelmente esses dados. No entanto, a idade, após a perturbação, teve um aumento na entropia, o que significa que houve mais variabilidade nos valores, tornando esses dados menos previsíveis. Entre as desvantagens dessa abordagem, destaca-se o fato de que, se o ruído for excessivamente pequeno, os dados ainda podem ser passíveis de reidentificação. Por outro lado, se o ruído for muito grande, as estatísticas podem ser distorcidas, o que pode afetar a análise de dados.

Os dados pessoais, como nome, CPF, e-mail e endereço, foram completamente removidos antes de aplicar a técnica de perturbação. Isso garante que esses dados não sejam preservados, mitigando os riscos de exposição. No caso dos dados numéricos, o ruído foi aplicado, mas, como alguns atributos estavam em formato de string, esses foram descartados, não sofrendo a perturbação diretamente. Em termos de impacto na privacidade, essa abordagem proporciona uma proteção forte, pois a remoção dos dados pessoais impede qualquer possibilidade de reidentificação direta. Contudo, isso também resulta na perda da possibilidade de vincular os dados anonimizados a outras bases de dados externas, limitando o uso de informações que possam ser associadas a outros contextos.

\subsection{K-Anonimato}
A técnica de K-Anonimato, que garante que cada grupo tenha pelo menos k registros idênticos, proporciona uma proteção adicional contra a reidentificação, mas resulta em uma perda moderada de informação. A entropia, apesar de não cair para 0\%, indica que os dados ainda mantêm variação útil. No entanto, a cidade perdeu 36,25\% da informação, o que implica uma redução na granularidade, mas ainda preserva uma boa diversidade. Já o estado civil e a cor sofreram uma perda de 32-39\% da informação, mas continuam mantendo alguma variabilidade. Entre as desvantagens dessa abordagem, se o valor de k for muito alto, pode haver uma generalização excessiva, resultando em uma perda de informações importantes. Além disso, encontrar o equilíbrio entre privacidade e utilidade pode ser desafiador, pois é preciso garantir que a proteção da identidade não prejudique a utilidade dos dados para análises.

Os dados pessoais foram completamente removidos antes de aplicar a técnica, o que elimina o risco de exposição direta desses dados. Além disso, os grupos foram formados de maneira que nenhuma pessoa possa ser identificada individualmente dentro de um grupo de pelo menos k registros idênticos. Isso garante uma proteção robusta à privacidade, tornando impossível a identificação de um único indivíduo dentro de cada grupo. No entanto, se o valor de k for pequeno, pode haver algum risco residual de reidentificação, especialmente quando esses dados são cruzados com outras bases de dados externas. Isso destaca a importância de ajustar k adequadamente para balancear a proteção à privacidade e a preservação da utilidade dos dados.

\subsection{Generalização}

A técnica de generalização, que substitui valores específicos por categorias amplas, resulta em uma alta perda de informação. A cidade, o estado e o problema de saúde sofreram uma perda de 100\% da variabilidade, sendo reduzidos a um único valor ou categoria. Em contrapartida, atributos como cor e religião perderam entre 39\% e 50\% da variabilidade, indicando que alguma diversidade foi preservada. Entre as desvantagens dessa abordagem, destaca-se a perda de padrões valiosos nos dados. Além disso, se a generalização for excessiva, os dados podem se tornar inúteis para análises detalhadas, pois a granularidade dos dados é perdida.

Os dados pessoais, como nome, CPF, e-mail e endereço, foram completamente removidos antes de aplicar a técnica, garantindo que não houvesse possibilidade de identificação direta. Outros dados, como a cidade, foram generalizados (por exemplo, cidade → região) para reduzir o risco de reidentificação. Embora essa generalização ofereça uma forte proteção à privacidade, ela resulta em uma perda de granularidade, o que pode afetar a utilidade dos dados para análises mais detalhadas. Mesmo assim, a abordagem garante uma segurança extrema, uma vez que os identificadores foram removidos e as demais informações foram suficientemente generalizadas para proteger contra ataques de reidentificação.

\section{Discussão sobre os Resultados e o Método Computacional para Anonimização}

A análise das diferentes técnicas de proteção de dados leva a conclusões importantes sobre a melhor escolha para alcançar diversos objetivos relacionados à privacidade e à utilidade dos dados. Se o foco principal for proteger dados sem comprometer a variabilidade das informações, as técnicas de perturbação ou K-Anonimato se destacam como as melhores opções. Ambas preservam a variabilidade dos dados, proporcionando um equilíbrio entre segurança e utilidade, tornando-as ideais para contextos onde a reidentificação precisa ser evitada, mas ainda se deseja manter padrões e detalhes importantes.

Por outro lado, quando o objetivo é garantir máxima privacidade, mas com a aceitação de uma perda significativa de informação, agregação e generalização são as técnicas mais seguras. Embora essas abordagens proporcionem uma proteção robusta contra ataques de reidentificação, elas resultam na destruição da variabilidade dos dados, o que pode comprometer sua utilidade para análises detalhadas. Portanto, esses métodos são mais indicados quando a privacidade é a prioridade, mesmo que se perca granularidade.

Se a necessidade for realizar uma análise estatística sem risco de reidentificação, a perturbação surge como uma boa opção. Ao adicionar ruído aos dados, a técnica mantém os padrões estatísticos relevantes enquanto protege as informações pessoais, garantindo que os dados possam ser usados para análises sem comprometer a privacidade.

A escolha do método mais adequado depende diretamente do objetivo final: segurança máxima ou um equilíbrio entre privacidade e utilidade. Quando o foco é a proteção dos dados pessoais, todas as técnicas examinadas removeram informações sensíveis, como nome, CPF, e-mail e endereço, o que é considerado a melhor prática para evitar reidentificação direta. Para aqueles que buscam garantir a máxima privacidade, agregação e generalização são as opções mais eficazes, enquanto para aqueles que necessitam preservar alguma utilidade estatística sem comprometer a segurança, K-Anonimato e perturbação são as escolhas mais eficientes.

É de fundamental importância compreender que a mera remoção de dados pessoais não garante a anonimização, em conformidade com a
LGPD. Evidências presentes na literatura científica demonstram que, mesmo em bases de dados públicas, a reversão dos dados foi possível. Portanto, para assegurar a conformidade com a LGPD, a aplicação de técnicas de anonimização torna-se uma estratégia essencial, especialmente em situações em que os dados são compartilhados com terceiros.

Além disso, é importante ressaltar que todos os resultados apresentados são provenientes de processo automatizado, que pode ser replicado para qualquer dataset. Neste processo o algoritmo calcula a entropia dos dados originais e anonimizados para avaliar a quantidade de informação preservada, comparando as métricas para verificar o impacto na utilidade dos dados. A análise da acurácia dos dados anonimizados em relação aos originais garante a fidelidade dos dados para análises subsequentes. A automação desse processo assegura avaliações rápidas, consistentes e objetivas, eliminando erros humanos e melhorando a precisão e confiabilidade. Medir a eficiência do processo, incluindo o tempo de execução, otimiza o desempenho, e o armazenamento das métricas facilita a documentação e transparência. Isso contribui para a qualidade do software e a conformidade com regulamentos de privacidade e segurança.

\section{Conclusão}

A proteção de dados pessoais é um aspecto essencial no desenvolvimento de software contemporâneo, especialmente à luz da LGPD e do GDPR. Estes regulamentos impõem exigências rigorosas que afetam significativamente a maneira como os softwares são projetados, desenvolvidos e mantidos, com foco na proteção da privacidade dos indivíduos. Este estudo investigou diversas técnicas de anonimização de dados, analisando suas efetividades na conformidade com a LGPD e demonstrando a importância de um equilíbrio entre privacidade e utilidade dos dados.

As técnicas de anonimização como agregação, generalização, perturbação e k-anonimato foram aplicadas a datasets com dados pessoais e sensíveis. Os resultados mostraram variações significativas na eficácia de cada método. A agregação e o k-anonimato se destacaram na preservação da privacidade, mas com um impacto substancial na utilidade dos dados. A generalização, embora eficaz em anonimização completa, resultou na perda total de utilidade dos dados, enquanto a perturbação apresentou um equilíbrio promissor, preservando a utilidade dos dados com uma perda de informação moderada.

A implementação de um algoritmo automatizado para testar a qualidade das técnicas de anonimização demonstrou ser um avanço significativo. Este algoritmo avalia a entropia e a acurácia dos dados anonimizados, proporcionando uma análise detalhada e objetiva da eficácia das técnicas. A automatização elimina a variabilidade humana e assegura consistência nos resultados, beneficiando diretamente a qualidade do software. Além disso, a medição da eficiência do processo e a documentação rigorosa das métricas de desempenho facilitam a conformidade com regulamentações de privacidade, promovendo um desenvolvimento de software mais seguro e confiável.

Este estudo contribui para uma compreensão mais profunda das práticas de anonimização e destaca a importância de abordagens automatizadas na manutenção da qualidade e conformidade do software, alinhando-se às exigências legais e éticas de proteção de dados pessoais.

\bibliographystyle{sbc}
\bibliography{sbc-template}

\end{document}